


\magnification\magstep1
\baselineskip14pt
\vsize24.0truecm

\input miniltx
\input graphicx


\font\bigbf=cmbx12
\font\csc=cmcsc10
\font\smallrm=cmr8
\def\hoppp{\bigskip\noindent}
\def\hopp{\medskip\noindent}
\def\hop{\smallskip\noindent}

\def\half{\hbox{$1\over2$}}

\def\section{\bigskip}
\def\subsection{\medskip}

\def\data{{\rm data}}

\def\midd{\,|\,}

\def\pred{{\rm pred}}
\def\res{{\rm res}}

\def\BIC{{\rm BIC}}

\def\newline{\hfill\par\vskip-1pt\noindent}

\def\fermat#1{\setbox0=\vtop{\hsize4.00pc
        \smallrm\raggedright\noindent\baselineskip9pt
        \rightskip=0.5pc plus 1.5pc #1}\leavevmode
        \vadjust{\dimen0=\dp0
        \kern-\ht0\hbox{\kern-4.00pc\box0}\kern-\dimen0}}


\font\cyr=wncyr10
\font\cyb=wncyb10
\font\bigcyb=wncyb10 at 12 pt

\font\cyi=wncyi10
\font\cyss=wncyss10
\font\csc=cmcsc10
\font\ssfont=cmss10
\font\bigbf=cmbx10 at 12 pt

\font\smallrm=cmr8
\font\smallcyr=wncyr8

\def\e3{\rm\"{\cyr e}}
\def\bolde3{\bf\"{\cyb e}}
\def\bigbolde3{\bigbf\"{\bigcyb e}}
\def\itae3{\it\"{\cyi e}}
\def\sse3{\ssfont\"{\cyss e}}
\def\smalle3{\smallrm\"{\smallcyr e}}
\def\j3{{\rm\u{\cyr i}}}
\def\boldj3{{\bf\u{\cyb i}}}
\def\bigboldj3{{\bigbf\u{\bigcyb i}}}
\def\itaj3{{\it\u{\cyi i}}}
\def\ssj3{{\ssfont\u{\cyss i}}}
\def\extraj3{{\csc\u{\cysc i}}}
\def\smallj3{{\smallrm\u{\smallcyr i}}}
\def\DD{{D$^*$}}
\def\TD{{\cyr TD}}
\def\Sh{{\cyr Sh}}
\def\Kr{{\cyr Kr}}
\def\tikhiidon{{Tikhi\u\i{} Don}}

\hopp
{\bigbf And Quiet Does Not Flow the Don:}

\hop     
{\bigbf Statistical Analysis of a Quarrel Between Nobel Prize Laureates} 

\hopp 
{\bf Nils Lid Hjort$^{1,2}$} 

\hop
{\bf $^1$Department of Mathematics, University of Oslo} 

\hop
{\bf $^2$Centre of Advanced Study, Norwegian Academy of Science and Letters}

\hop
{\bf September 2006} 

\hoppp
The Nobel Prize in literature 1965 was awarded
Mikhail Sholokhov (1905--1984), for the epic
novel {\cyi Tihi\itaj3{} Don} about Cossack life and
the birth of a new Soviet society
({\sl And Quiet Flows the Don}, 
or {\sl The Quiet Don}, in different translations). 
Sholokhov has been compared to Tolsto\u\i{} and was
at least a generation ago called `the greatest of our writers'
in the Soviet Union. In Russia alone his books 
have been published in more than a thousand editions, 
selling in total more than sixty million copies. 
He was an elected member of the USSR Supreme Soviet, 
the USSR Academy of Sciences, 
and of the CPSU Central Committee. 

But in the autumn of 1974 an article was
published in Paris, {\cyi Stremya `Tihogo Dona' (Zagadki romana)}
{\sl (`The Rapids of Quiet Don: the Enigmas of the Novel')}, 
by the author and critic \DD. He claimed that
\tikhiidon{} was not at all Sholokhov's work,
but that it rather was written by
Fiodor Kriukov, a more obscure author who fought
against bolshevism and died in 1920. The article was given
credibility and prestige by none other than
Aleksandr Solzhenitsyn (a Nobel prize winner five years after Sholokhov), 
who wrote a preface giving full support to \DD's conclusion. 
Scandals followed, also touching the upper echelons of Soviet society,
and Sholokhov's reputation was faltering abroad 
(see e.g.~Doris Lessing's (1997) comments; 
`vibrations of dislike instantly flowed between us'). 
Are we in fact faced with one of the most flagrant cases 
of plagiarism in the history of literature? 

\hoppp
{\bf Approaching disputed authorship cases}

\hop
The first reaction to accusations of plagiarism 
or to cases of disputed authorship is 
perhaps simply to listen to the points being made,
checking the strength of argumentation by common sense
or if required with the careful skepticism of a court of law.
In the present case the claims made would perhaps 
be classified as unsubstantial. Various rumours 
were in circulation already from 1930, as detailed
in Kjetsaa et al.'s (1984) account. Solzhenitsyn's (1974) 
preface appears to rest on the opinions that 
(i) such a young and relatively un-educated person 
could not produce so much high literature in such 
a short time-span;  
(ii) all his later work was produced at a much slower pace,
and has a lower literary quality; 
(iii) Kriukov's background and publications 
(summarised in Solzhenitsyn's own afterword to the 1974 publication) 
fit the story teller's perspectives better. 
To this was also added the unfortunate fact that 
Sholokhov's personal author's archives could not be found. 
Further elaborations, partly also qua attempts at 
linguistic and stylistic analyses, can be seen 
in \DD{} (1974). 

If the case still warrants further discussion, after 
initial scrutiny, one may enter the intriguing but 
difficult terrain of sorting out an author's 
or artist's `personal style', whether in stylistic expression, 
or via smaller idiosyncrasies, or perhaps a bit
more grandly in their themes and how these are developed. 
In a famous essay, Sir Isaiah Berlin (1953) 
made a bold attempt at sorting Russian authors into 
`hedgehogs' and `foxes', after the old Greek saying
that Erasmus Rotterdamus records as 
{\it `multa novit vulpes, verum echinus unum magnum'}: 
the fox knows many tricks, but the hedgehog masters
one big thing. Thus Dostoyevski\u\i{} and Ibsen are hedgehogs
while Pushkin and Tolsto\u\i{} were foxes -- the latter trying
however very hard to become a hedgehog, according to Berlin. 
In Hjort (2006) I follow such a 
literary classification challenge by arguing, 
in three languages, that Carl Barks is a fox while
Don Rosa is a hedgehog. 
See in this connection also Gould (2003) who uses 
the hedgehog vs.~fox dichotomy to address the
misconceived gap between sciences and the humanities
(in the best spirit of the Centre of Advanced Studies).  

\includegraphics[scale=0.33]{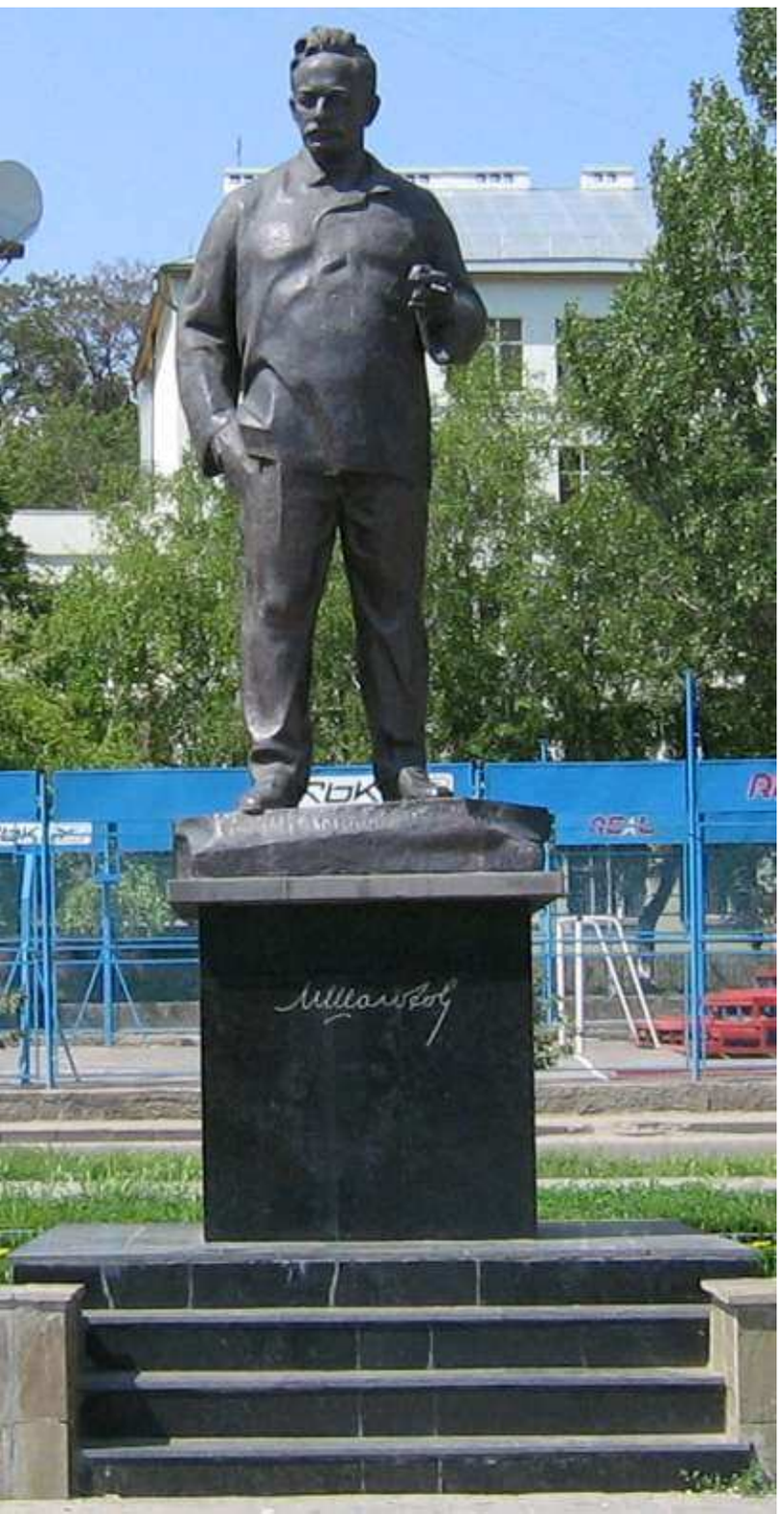}
\includegraphics[scale=0.16]{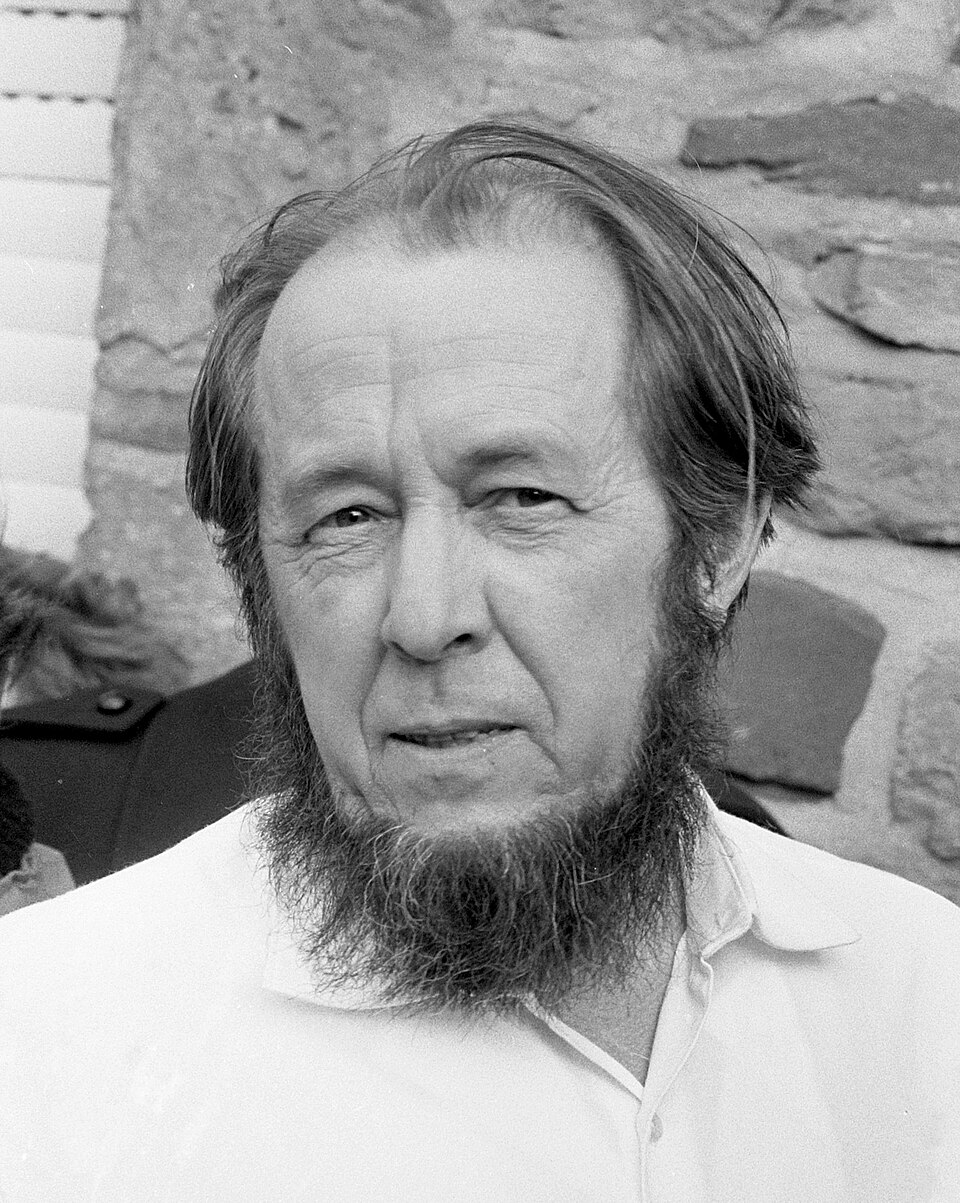}


\vskip0.1truecm
{\medskip\sl\narrower\noindent\baselineskip11pt
{\csc Figure 1:}
Mikhail Sholokov (1905--1984), Nobel Laureate 1965; 
Aleksandr Sol\-zhenitsyn (1918--), Nobel Laureate 1970,
who wishes to put the 1965 winner down from his pidestal. 
\smallskip}

But even experts on literature, art and music are prone
to making occasional mistakes, as demonstrated often enough,
and it is clear that independent arguments based on
quantitative comparisons are of interest -- if not 
taken as `direct proof', then such comparisons
may at least offer independent objective evidence 
and sometimes additional insights.
In such a spirit an inter-Nordic research team was formed 
in the course of 1975, captained by Geir Kjetsaa,
a professor of Russian literature
at the Department of Literature, Regional Studies
and European Languages at the University of Oslo,
with the aim of disentangling the Don mystery.
In addition to various linguistic analyses and some
doses of detective work, quantitative data were 
gathered and organised, 
for example relating to word lengths, 
frequencies of certain words and phrases, 
sentence lengths, grammatical characteristics, etc.
These data were extracted from three corpora:
(i) \Sh, or Sh, from published work guaranteed to be by Sholokhov;
(ii) \Kr, or Kr, that which with equal trustworthiness
came from the hand of the alternative hypothesis Kriukov; and
(iii) \TD, or TD, the Nobel winning apple of discord. 
Each of the corpora has about 50,000 words.
My contribution here is to squeeze clearer author discrimination
and some deeper statistical insights 
out of some of Kjetsaa et al.'s data. 

\hoppp
{\bf Sentence length distributions}

\hop
I shall here focus on the statistical distribution
of the number of words used in sentences, as a
possible discriminant between writing styles.
Table 1, where the first five columns have been compiled via other tables 
in Kjetsaa et al.~(1984), summarises these data,
giving the number of sentences in each
corpus with lengths between 1 and 5 words,
between 6 and 10 words, etc. The sentence length
distributions are also portrayed in
Figure 2, along with
fitted curves that are described below.
The statistical challenge is to explore whether
there are any sufficiently noteworthy differences
between the three empirical distributions, and, if so,
whether it is the upper or lower distribution
of Figure 2 that most resembles 
the one in the middle.

{\smallskip\narrower\noindent\baselineskip12pt
{\csc Table 1:}
{\sl \tikhiidon: number of sentences $N_x$ in the three
corpora Sh, Kr, TD of the given lengths,
along with predicted numbers $\pred_x$ 
under the four-parameter model (1), 
and Pearson residuals $\res_x$,
for length groups $x=1,2,3,\ldots,13$. 
The average sentence lengths are 12.30, 13.12, 12.67
for the three corpora, and the variance to mean 
dispersion ratios are 6.31, 6.32, 6.57.
\smallskip}}

{{\smallskip\obeylines\tt\baselineskip11pt
~~~~~~~~~~~~~observed:~~~~~~~~~~~~predicted:~~~~~~~~~~~~residuals:
from~to~~~Sh~~~~Kr~~~~TD~~~~~Sh~~~~~~Kr~~~~~~TD~~~~~~Sh~~~~~Kr~~~~TD~~
~~1~~~5~~~799~~~714~~~684~~~803.4~~~717.6~~~690.1~~-0.15~~-0.13~~-0.23~
~~6~~10~~1408~~1046~~1212~~1397.0~~1038.9~~1188.5~~~0.30~~~0.22~~~0.68~
~11~~15~~~875~~~787~~~826~~~884.8~~~793.3~~~854.4~~-0.33~~-0.22~~-0.97~
~16~~20~~~492~~~528~~~480~~~461.3~~~504.5~~~418.7~~~1.43~~~1.04~~~3.00~
~21~~25~~~285~~~317~~~244~~~275.9~~~305.2~~~248.1~~~0.55~~~0.67~~-0.26~
~26~~30~~~144~~~165~~~121~~~161.5~~~174.8~~~151.1~~-1.38~~-0.74~~-2.45~
~31~~35~~~~78~~~~78~~~~75~~~~91.3~~~~96.1~~~~89.7~~-1.40~~-1.85~~-1.55~
~36~~40~~~~37~~~~44~~~~48~~~~50.3~~~~51.3~~~~52.1~~-1.88~~-1.02~~-0.56~
~41~~45~~~~32~~~~28~~~~31~~~~27.2~~~~26.8~~~~29.8~~~0.92~~~0.24~~~0.23~
~46~~50~~~~13~~~~11~~~~16~~~~14.5~~~~13.7~~~~16.8~~-0.39~~-0.73~~-0.19~
~51~~55~~~~~8~~~~~8~~~~12~~~~~7.6~~~~~6.9~~~~~9.4~~~0.14~~~0.41~~~0.85~
~56~~60~~~~~8~~~~~5~~~~~3~~~~~4.0~~~~~3.5~~~~~5.2~~~2.03~~~0.83~~-0.96~
~61~~65~~~~~4~~~~~5~~~~~8~~~~~2.1~~~~~1.7~~~~~2.9~~~1.36~~~2.51~~~3.04~
~~~~~~~~~4183~~3736~~3760~      
\smallskip}}

A very simple model for sentence lengths is that
of the Poisson, but one sees quickly that the
variance is larger than the mean (in fact with
a factor of around six, see Table 1). 
Another possibility is that of a mixed Poisson, 
where the parameter is not constant 
but varies in the world of sentences.
If $Y$ given $\lambda$ is Poisson with this parameter,
but $\lambda$ has a Gamma $(a,b)$ distribution,
then the marginal takes the form
$$f^*(y,a,b)={b^a\over \Gamma(a)}{1\over y!}
  {\Gamma(a+y)\over (b+1)^{a+y}}
  \quad {\rm for\ }y=0,1,2,\ldots, $$
which is the negative binomial. Its mean is $\mu=a/b$
and its variance $a/b+a/b^2=\mu(1+1/b)$, indicating
the level of over-dispersion.
Fitting this two-parameter model to the data was also
found to be too simplistic; clearly the muses had
inspired the novelists to transform their passions into patterns
more variegated than those dictated by a mere negative binomial,
their artistic outpourings also appearing to display
the presence of two types of sentences, the rather long ones
and the rather short ones, spurring in turn
the present author on to the following mixture of one Poisson,
that is to say, a degenerate negative binomial,
and another negative binomial, with a modification stemming
from the fact that sentences containing zero words
do not really count among Nobel literature laureates
(with the notable exception of a 1958 story by Heinrich B{\"o}ll):
$$ f(y,p,\xi,a,b)=p{\exp(-\xi)\xi^y/y!\over 1-\exp(-\xi)}
   +(1-p){f^*(y,a,b)\over 1-f^*(0,a,b)} \eqno(1)$$
for $y=1,2,3,\ldots$. It is this four-parameter family
that has been fitted to the data in Figure~2. 
The model fit is judged adequate, see Table 1, which in addition 
to the observed number $N_x$ shows the expected 
or predicted number $\pred_x$ of sentences
of the various lengths, for length groups $x=1,\ldots,13$.  
Also included are Pearson residuals $(N_x-\pred_x)/\pred_x^{1/2}$. 
These residuals should essentially
be on the standard normal scale if the parametric model
used to produce the predicted numbers is correct; here
there are no clear clashes with this hypothesis,
particularly in view of the large sample sizes
involved, with respectively 4183, 3736, 3760 sentences
in the three corpora. The $\pred_x$ numbers in the table
stem from minimum chi squared fitting for each of the
three corpora, i.e.~finding parameter estimates to minimise
$\sum_x\{N_x-\pred_x(\theta)\}^2/\pred_x(\theta)^2$ 
with respect to the four parameters, where
$\pred_x(\theta)=np_x(\theta)$ in terms of the
sample size for the corpus worked with and
the inferred probability $p_x(\theta)$ of writing
a sentence with length landing in group~$x$.

\hoppp
{\bf Statistical discrimination and recognition}

\hop
Kjetsaa's group quite sensibly put up Sholokhov's
authorship as the null hypothesis, and \DD's speculations
as the alternative hypothesis, in several of their analyses.
Here I shall formulate the problem in terms of selecting
one of three models, inside the framework of three data sets
from the four-parameter family (1):

\smallskip 
\item{$M_1$:} 
Sholokhov is the rightful author,
so that text corpora Sh and TD come from the same
statistical distribution, while Kr represents another;

\item{$M_2$;} 
\DD{} and Solzhenitsyn were correct in
denouncing Sholokhov, whose text corpus Sh is
therefore not statistically compatible with Kr and TD,
which are however coming from the same distribution; and

\item{$M_0$:} 
Sh, Kr, TD represent three statistically
disparate corpora.
\smallskip 

\noindent 
Selecting one of these models via statistical methodology 
will provide an answer to the question of who is 
most probably the author. 

\smallskip
\centerline{\includegraphics[scale=0.40,angle=270]{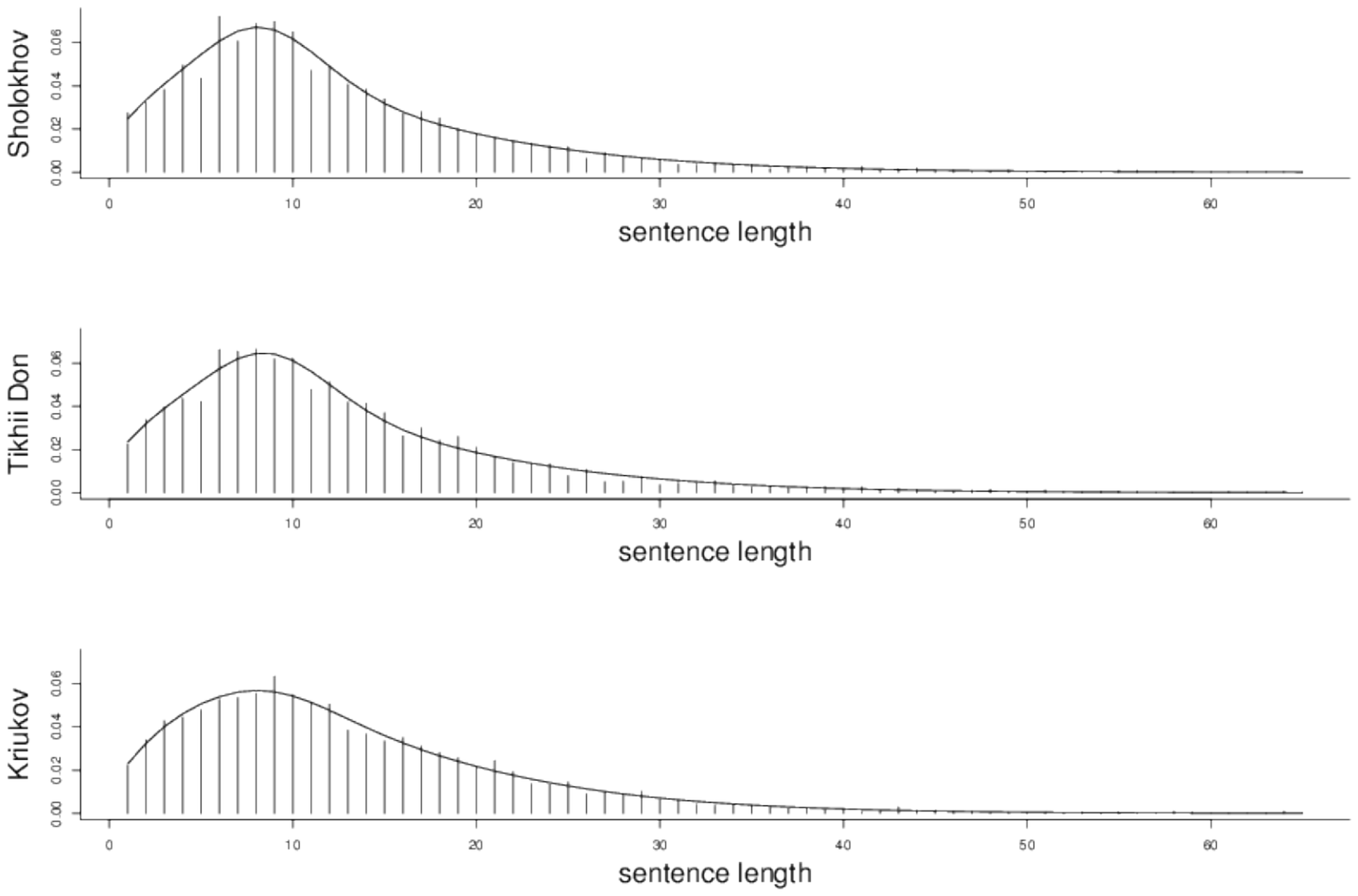}} 

\smallskip
\vskip0.1truecm
{\medskip\sl\narrower\noindent\baselineskip11pt
{\csc Figure 2.}
Sentence length distributions, from 1 word to 65 words,
for Sholokhov (top), Kriukov (bottom), and for
`The Quiet Don' (middle). Also shown, as continuous
curves, are the distributions (1),
fitted via maximum likelihood. 
The parameter estimates for $(p,\xi,a,b)$ are 
$(0.18,0.10,2.09,0.16)$ for Sh,
$(0.06,9.84,2.24,\allowbreak 0.18)$ for Kr, and
$(0.17,9.45,2.11,0.16)$ for TD. 
\smallskip}

Various model selection methods may now be applied,
to assist in ranking models $M_1$, $M_2$, $M_0$ 
by plausibility; 
see Claeskens and Hjort (2007) for a broad overview.  
Here I choose to concentrate on an approach that
hinges on the precise evaluation of a posteriori probabilities 
of the three models, given all available data, 
having started with any given a priori probabilities.
This makes it possible for different experts,
with differing degrees of prior opinions as to 
who is the most likely author, to revise their 
model probabilities in a coherent manner. 
This is quite similar to the so-called Bayesian Information Criterion
(BIC), but certain aspects of the present situation 
calls for refinements that make the analysis
reported on here more precise than the traditional BIC. 
These refined scores are called $\BIC^*$ below; 
models with larger scores are more probable
than those with smaller scores. 
\eject 

Let in general $p(M_1)$, $p(M_2)$, $p(M_0)$
be any engaged prior probabilities for the
three possibilities; we should perhaps take $p(M_0)$
close to zero, for example. Solzhenitsyn would 
take $p(M_1)$ rather low and $p(M_2)$ rather high,
while more neutral observers would perhaps start
with these two being equal to $\half$ and $\half$. 
Writing next $\theta_1$, $\theta_2$, $\theta_3$
for the three parameter vectors $(p,\xi,a,b)$,
for respectively Sh, Kr, TD, model $M_1$ holds that
$\theta_1=\theta_3$ while $\theta_2$ is different;
model $M_2$ claims that $\theta_2=\theta_3$ while
$\theta_1$ is different; and finally model $M_0$
is open for the possibility that the three parameter
vectors must be declared different. 
The posterior model probabilities may be computed as 
$$p(M_j\midd\data)=
  {p(M_j)\exp(\half\BIC_j^*)\over 
  p(M_0)\exp(\half\BIC_0^*)+
  p(M_1)\exp(\half\BIC_1^*)+
  p(M_2)\exp(\half\BIC_2^*)} \eqno(2)$$
for $j=0,1,2$. Space does not allow explaining
in detail here how the three $\BIC^*$ scores 
are computed, but they involve finding maximum likelihood
estimates of parameters under the three models 
and their precision matrices, along with 
accurate approximations of various integrals of dimension 8 and 12; 
see Claeskens and Hjort (2007, Section 5.4)
for a more detailed exposition of the somewhat 
elaborate mathematics involved. 

\hoppp
{\bf Conclusions}

\hop
Using (2) to compute posterior model probabilities
yields numbers close to zero for $M_2$ and $M_0$ 
and very close to one for $M_1$. With equal 
prior probabilities we actually find 0.998 for Sholokhov 
with the remaining 0.002 shared between Kriukov 
and the neutral model that the three corpora are disparate, 
and even Solzhenitsyn, starting with perhaps 
$p(M_1)=0.05$ and $p(M_2)=0.95$, 
will be forced to revise his $M_1$-probability to 0.99
and down-scale his $M_2$-probability to 0.01. 
We may conclude that the sentence length data speak
very strongly in Sholokhov's favour, 
and dismiss \DD's allegations as speculations:
the Stalin Prize in Literature 1941 went to the right person. 
These figures might sound surprisingly clear-cut, 
in view of the relative similarity of the distributions
portrayed in Figure 2.
The reason lies with the large sample sizes, which
increases detection power. 

Of course the probabilities that come from applying (2)
depend on the precise nature of (1),
and must be interpreted with some caution. 
There would be other parametric families that also
would fit the data, like a five-parametric mixture 
of two negative binomials, and these would lead 
to similar but not fully identical posterior model
probabilities. 

There have been various statistical studies
related to disputed authorship cases in the literature,
and one would expect this branch to expand with 
the increasing ease with which large texts may be 
inspected and analysed stylometrically via computers.  
Cox and Brandwood (1959) applied discriminant analysis
in an attempt to order Plato's dialogues chronologically, 
using stress patterns of the last five syllables 
in each sentence. 
Mosteller and Wallace (1964, 1984) examined 
the federalist papers, where 12 essays were published
anonymously, and used stylometry to demonstrate
that James Madison is the likely author. 
Thisted and Efron (1987) used empirical Bayes methods 
to assess whether a poem discovered in the 1980ies 
could be attributed to Shakespeare.

It should be pointed out 
that some statistical techniques might be quite
successful at finding differences between authors'
or artists' styles, via qualitative measurements
and analyses thereof, even if such discrimination 
abilities might not offer real insights per se 
as to the `real processes' that produce `real art'.
Thus Lyu, Rockmore and Farid (2004) provide digital
techniques for distinguishing Pieter Brueghel the elder
from various impostors, without pretensions of quantifying
what makes Brueghel the elder `better' than the others.  
I believe methods of the type discussed in the 
present article may be used in other studies 
involving `authors' fingerprints',
and that some such could throw light also on the 
intrinsically artistic aspects of the works under study.
It is my ambition to demonstrate statistically  
that Bach cannot be the composer of Bach Cantata \#189,
for example, as {\it Meine Seele r\"uhmt und preist},
although beautiful, is `too plain' in parameters 
pertaining to musical variation to be the real thing. 
For the Don case, analysis based on model (1) 
shows that Kriukov has a significantly smaller $p$ value
than has Sholokhov (while other parameters match 
reasonably well, see Figure 2), which a bit speculatively is 
an indication that the latter varied the form 
of his sentences more than the former. 

\hoppp
{\bf References}

\def\ref#1{{\noindent\hangafter=1\hangindent=20pt
  #1\smallskip}}
\parindent0pt
\baselineskip12pt
\parskip3pt
\smallskip

\ref{%
Berlin, I. (1953). 
{\sl The Hedgehog and the Fox.}
Weidenfeld \& Nicolson, London.}

\ref{%
Claeskens, G.~and Hjort, N.L. (2007).
{\sl Model Selection and Model Averaging.}
Cambridge University Press, Cambridge.}

\ref{%
Cox, D.R.~and Brandwood, L. (1959):
On a discriminatory problem connected with the works of Plato.
{\sl Journal of the Royal Statistical Society} {\bf B 21}, 195--200.}

\ref{%
\DD{} (1974).
{\cyi Stremya `Tihogo Dona' (Zagadki romana)}
{\sl (The Rapids of Quiet Don: the Enigmas of the Novel)}.
YMCA-Press, Paris.
(The full Russian text is available at  
{\tt newchrono.ru/frame1/Literature/QuietDon/stremya.htm}.)}


\ref{%
Gould, S.J. (2003).
{\sl The Hedgehog, the Fox, and the Magister's Pox:
Mending and Minding the Misconceived Gap
Between Science and the Humanities.}
Vintage, London.}

\ref{%
Hjort, N.L. (2006).
The fox, the hedgehog, and the ducks. 
Preface to {\sl Hall of Fame: Don Rosa Volume III}
(in Norwegian, Danish, Swedish, in separate editions).
Egmont, K\o benhavn.}

\ref{%
Kjetsaa, G., Gustavsson, S., Beckman, B.~and Gil, S. (1984). 
{\sl The Authorship of The Quiet Don.}
Solum/Humanities Press, Oslo.}

\ref{%
Lessing, D. (1997).
{\sl Walking in the Shade: Volume Two of My Autobiography 1949 to 1962.}}

\ref{%
Lyu, S., Rockmore, D.~and Farid, H. (2004).
A digital technique for art authentication.
Proceedings of the National Academy of Sciences 
of the U.S.A.~{\bf 101}, 17006--17010.}

\ref{%
Mosteller, F.~and Wallace, D. (1984).
{\sl Applied Bayesian and Classical Inference: 
The Case of the Federalist Papers.}
Springer-Verlag, New York.
[Extended edition of their 1964 book,  
{\sl Inference and Disputed Authorship: The Federalist}, 
Addison-Wesley, Reading, Massachusetts.]}

\ref{%
Solzhenitsyn, A.I. (1974).
{\cyi Nevyrvannaya ta{\itaj3}na}
{\sl (The not yet uprooted secret)}.
Preface to \DD{} (op.cit.).}

\ref{%
Thisted, R.~and Efron, B. (1987).
Did Shakespeare write a newly discovered poem?
{\sl Biometrika} {\bf 74}, 445--455.}

\bye